\documentstyle[preprint,prl,aps]{revtex}

\tightenlines

\begin{document}
\title{Fokker-Planck description and diffusive phonon heat transport}
\author{Kwok Sau Fa}
\address{Departamento de F\'{\i}sica, Universidade Estadual\\
de Maring\'a, Av. Colombo 5790, 87020-900 Maring\'a-PR, Brazil}
\maketitle

\begin{abstract}
We propose a prescription based on the Fokker-Planck equation in the
Stratonovich approach, with the diffusion coefficient dependent on temporal
and spatial coordinates, for describing heat conduction by phonons in small
structures. This equation can be analytically solved for a broad class of
diffusion coefficients. It can also describe non-Gaussian processes.
Further, it generalizes the model investigated by Naqvi and Waldenstr$\phi $%
m (PRL, 95 (2005), 065901). We show that our solutions can fit well the
results derived from the Boltzmann equation.

\newpage
\end{abstract}

The processes of the heat conduction in physical systems involve the
microscopic transports of the energy carriers. In general, these processes
can be demarcated by certain characteristic time and length scales of the
energy carriers such as collision time, mean free time, relaxation time,
diffusion time, mean free path, relaxation length and diffusion length.
However, the descriptions of the heat conduction processes have encompassed
several theoretical approaches. The simplest one is based on the Fourier law 
\cite{joseph}. This law enjoys some universality in the description of the
heat transfer due to the fact that it can be applied to a wide range of the
physical systems with different energy carriers, in the macroscales. The
Fourier law breaks down for anomalous heat conduction systems, where the
thermal conductivity $\kappa $\ diverges with the system size $L$\ as $%
L^{\beta }$. Theoretical investigation based on a connection between the
anomalous diffusion processes and anomalous heat conduction in a
one-dimensional systems has been carried out \cite{li}. Also, the Fourier
law breaks down in the domain of the microscales such as the heat transport
in a thin film \cite{joshi}. For this last case, the equation of phonon
radiative transfer (EPRT) can describe well, however, it is difficult to be
solved. Recently, several authors have attempted to replace easier models
from which can give a good approximation of the EPRT results \cite
{chen,chen2,razi}. The ballistic-diffusive equations (BDE), derived from the
Boltzmann equation under the relaxation time approximation, which has been
introduced by Chen \cite{chen,chen2} can capture the behaviors of the
temperature and heat flux of the EPRT. An other description is based on the
Brownian motion which has been introduced by Naqvi and Waldenstr$\phi $m
(NW) \cite{razi}. This last model can describe the temperature of the EPRT
very well, but its heat flux deviates visibly in some spatial range (see
Fig. 1),where $t^{\ast }=t/\tau $, $\xi =x/L$, $\Delta T=T_{1}-T_{0}$, $%
\theta =(T-T_{0})/\Delta T$\ , $\phi =q/(Cv\Delta T)$,\ $\tau =l/v$ is the
mean-free time, $v$\ is the average of sound, $l$\ is the mean-free path and 
$Kn=l/L$\ is the Knudsen number.

In this letter, we propose to generalize the NW model. Our model is based on
the Fokker-Planck equation in the Stratonovich approach \cite{risken} with
the diffusion coefficient that depends on time and space. This model can be
analytically solved for a broad class of diffusion coefficients, and it also
presents interesting asymptotic properties \cite{kwok}. We show that our
solutions can give a good approximation of the EPRT results.

\ In order to motivate our proposal, we first discuss the NW model which is
given by

\begin{equation}
\partial _{t}T(x,t)=a(t)\partial _{xx}T(x,t)-b(t)\partial _{x}T(x,t)\text{ \
,}  \label{eq1}
\end{equation}
where $T(x,t)$\ is the temperature. In their analysis, Naqvi and \ Waldenstr$%
\phi $m have chosen $a(t^{\ast })=\kappa (1-e^{-t^{\ast }})$\ and $b(t)=0$,
where $\kappa =vl/3$\ is the thermal diffusivity. In this case, the heat
flux is given by $q=-(\lambda /\kappa )a(t^{\ast })\partial _{x}T(x,t)$,
where $\lambda =Cvl/3$ and $C$\ is the specific heat per unit volume. For $%
t^{\ast }\gg 1$, \ the model recovers the Fourier equation. As have been
noted by the authors, the model (\ref{eq1}) describes a Gaussian process and
it can only approximate to the results of EPRT. For $b(t)=0,$\ we can show
that the solutions of Eq. (\ref{eq1}) can not \ be improved for any choice
of $a(t)$, and it can only obtain the similar results described by $%
a(t^{\ast })=\kappa (1-e^{-t^{\ast }}).$\ In order to show this fact, we
plot $\Lambda \equiv -\phi /(\partial _{\xi }\theta )$\ against the
nondimensional coordinate $\xi $\ (Fig.2) from the data of EPRT (Fig.1). We
see that $\Lambda $\ \ does not remain constant. From the NW model we obtain 
$\Lambda =Kn(1-e^{-1})/3\cong 0.21$. This value approximates to the first
part of the curve of Fig.2 well. Then, around the value $\xi =0.6$, the
curve begins to deviate visibly from the value $0.21$. As can also be seen
from Fig. 1 the heat flux begins to deviate from the EPRT result around the
value $\xi =0.6$. This shows that the model (\ref{eq1}), for $b(t)=0$, can
not be improved. On the other hand, for any function of $b(t)$\ different
from $a(t)$, the solution of equation (\ref{eq1}) may not be easily obtained
and the method of separation of variables can not be used, either.

In our proposal we consider the following equation

\begin{equation}
\partial _{t}T(x,t)=\kappa a(t^{\ast })\partial _{x}\left[ D(x)\partial
_{x}\left( D(x)T(x,t)\right) \right] \text{ \ }  \label{eq2}
\end{equation}
and the heat flux given by

\begin{equation}
q=-\lambda a(t^{\ast })D(x)\partial _{x}\left( D(x)T(x,t)\right) \text{ .}
\label{eq3}
\end{equation}
We note that if $T(x,t)$ is replaced by the probability density, then Eq. (%
\ref{eq2}) becomes a stochastic equation namely the Fokker-Planck equation
in the Stratonovich approach which is obtained from the Langevin equation
with a multiplicative noise term \cite{risken}. We see that Eq. (\ref{eq2})
generalizes the NW model with $b(t)=0$. In this case, we recover Eq. (\ref
{eq1}) for constant $D(x)$.

The application of Eqs. (2) and (3) to the physical systems is to consider a
slab of thickness $L$\ coupled to two thermal reservoirs. At time $t=0$, one
face (at $x=L)$\ is maintained at the temperature $T_{0}$, whereas the other
face (at $x=0$) is raised to a temperature $T_{1}$. Moreover, initially the
slab is maintained at a uniform temperature $T_{0}$. In order to compare
with the results of other approaches we use the nondimensional quantities
defined above. We also consider the diffusion coefficient only depends on
the nondimensional variables. The solution of Eq. (\ref{eq2}) in terms of
the nondimensional variables can be obtained by the transformations: d$u/$d$%
\xi =1/D(\xi )$ and d$s/$d$t^{\ast }=a(t^{\ast })$. Then, Eq. (\ref{eq2})
reduces to $\partial _{s}\rho (\xi ,s)=(\kappa \tau /L^{2})\partial
_{u}^{2}\rho (u,s)$,\ where $\rho =D\theta $. For convenience, we set $u(\xi
=0)=0$ and $s(t^{\ast }=0)=0$; \ and the solutions for $\theta (\xi ,t^{\ast
})$ and $\phi (\xi ,t^{\ast })$ subject to the boundary conditions above are
given by

\begin{equation}
\theta (\xi ,t^{\ast })=\frac{D_{0}}{D(\xi )}\left\{ 1-\frac{u(\xi )}{u_{1}}-%
\frac{2}{\pi }\sum_{m=1}^{\infty }\frac{\sin \left( \frac{m\pi u\left( \xi
\right) }{u_{1}}\right) \exp \left( -\frac{Kn^{2}(m\pi )^{2}s\left( t^{\ast
}\right) }{3u_{1}^{2}}\right) }{m}\right\}  \label{eq4}
\end{equation}
and \ 

\begin{equation}
\phi (\xi ,t^{\ast })=\frac{KnD_{0}a(t^{\ast })}{3u_{1}}\left\{
1+2\sum_{m=1}^{\infty }\cos \left( \frac{m\pi u\left( \xi \right) }{u_{1}}%
\right) \exp \left( -\frac{Kn^{2}(m\pi )^{2}s\left( t^{\ast }\right) }{%
3u_{1}^{2}}\right) \right\} \text{ ,}  \label{eq5}
\end{equation}
where $D_{0}$ is the value of $D(\xi )$\ at $\xi =0$\ and $u_{1}$ is the
value of $u(\xi )$ at $\xi =1$. For $D=1$, we recover the results of NW model

\begin{equation}
\theta (\xi ,t^{\ast })=\left\{ 1-\xi -\frac{2}{\pi }\sum_{m=1}^{\infty }%
\frac{\sin \left( m\pi \xi \right) \exp \left( -\frac{Kn^{2}(m\pi
)^{2}s\left( t^{\ast }\right) }{3}\right) }{m}\right\}  \label{eq6}
\end{equation}
and

\begin{equation}
\phi (\xi ,t^{\ast })=\frac{Kna(t^{\ast })}{3}\left\{ 1+2\sum_{m=1}^{\infty
}\cos \left( m\pi \xi \right) \exp \left( -\frac{Kn^{2}(m\pi )^{2}s\left(
t^{\ast }\right) }{3}\right) \right\} \text{ .}  \label{eq7}
\end{equation}

For our numerical investigation we choose $a(t^{\ast })=$\ $\kappa
(1-e^{-ht^{\ast }})$\ and $D(\xi )=p_{1}\left( 1+p_{2}\xi ^{n}\right)
/\left( 1+p_{3}\xi ^{n}\right) $, where $h,$ $p_{1},$ $p_{2}$\ and $p_{3}$\
are the parameters to be adjusted. For simplicity, we have investigated the
numerical solutions by using $n$ as an integer. It seems that the numerical
results, for $n=4,$\ are better than other values of $n$. In Fig. 3 and Fig.
4, we show the temperature and heat flux \ distributions obtained from Eqs. (%
\ref{eq4}) and (\ref{eq5}) by using $D_{1}(\xi )=\left( 1+p_{2}\xi
^{4}\right) /\left( 1+p_{3}\xi ^{4}\right) $ and $D_{2}(\xi )=p_{3}\left(
1+p_{2}\xi ^{4}\right) /[p_{2}\left( 1+p_{3}\xi ^{4}\right) ]$, respectively$%
.$\ Both the results can fit the EPRT results well. We see that our
numerical solutions, by using $D_{1}$,\ can fit the results of EPRT better
than those obtained by $D_{2}$. However, the advantage of $D_{2}$ is that it
tends to unity for $\xi \gg 1$. As have been noted by Joshi and Majumdar 
\cite{joshi} , when the size of a slab is much larger than the phonon mean
free path the heat transport can be modeled by the Fourier law.

In Fig. 5 we compare the temporal behaviors of the nondimensional heat flux
(at $\xi =0)$\ obtained from the BME and FPE.\ We note that the EPRT data
have not been included in Fig. 5 due to the fact that they have not been
rescaled \cite{razi}. The upper curves ($Kn=10$) show that our result has a
similar behavior of that described by the BME, but our curve is lower than
that described by the BME for not too small $t^{\ast }$.\ The middle curves (%
$Kn=1$) also show that our curve is lower than that of the BME for $t^{\ast
}>1$. We note that the BME curve has a pronounced minimum around the value $%
t^{\ast }=1.8$, whereas our curve decays to a lower point and then keep it
nearly straight. As a consequence, our curve can qualitatively reproduce the
behavior of the EPRT result (see Fig. 2 of Ref. \cite{chen}) better than
that of the BME. Finally, the lower curves show that our curve reproduces
the BME result.

As have been demonstrated in \cite{joshi,chen,razi} the Fourier equation and
Cattaneo equation \cite{catta} can not describe the EPRT results and
consequently they fail to describe the heat conduction on small scales such
as in a thin film. Moreover, the Fourier equation leads to a divergent heat
flux for $t^{\ast }\rightarrow 0$ and the Cattaneo equation produces
artificial heat flux oscillation. In this work we have mainly concentrated
our studies on the MBE and FPE, and their results have been compared with
those calculated by using the EPRT and BDE. Our results based on the FPE can
describe well the EPRT results. We have chosen the FPE in the Stratonovich
approach due to the fact that it can be analytically solved for a broad
class of diffusion coefficients, whereas in other approaches such as Ito and
postpoint discretization approaches (see \cite{kwok} and the references
therein) we do not have the same facility. However, there is no reason to
choose solely the FPE in the Stratonovich approach. In fact, numerical
calculation can be used to obtain the solutions of other approaches. In
order to choose which of the FPE approaches is more adequate for describing
the heat conduction on small scales we need further information of the
microscopic structure of the systems. We note that the FPE with the
diffusion coefficient that depends on time and space can describe
non-Gaussian processes with white noise \cite{kwok}. In particular, the
non-Gaussian processes obtained from the FPE are due to the dependence of
the diffusion coefficient on the spatial coordinate. Therefore, Eq. (\ref
{eq1}) can only describe Gaussian processes and it is a particular case of
Eq. (\ref{eq2}). Finally, we would mention that we have used the
coefficients $a(t^{\ast })=$\ $(1-e^{-ht^{\ast }})$\ and $D(\xi
)=p_{1}\left( 1+p_{2}\xi ^{n}\right) /\left( 1+p_{3}\xi ^{n}\right) $
because they are simple expressions and they can fit well the EPRT results.
However, other more elaborate expressions may also be employed for improving
our results above.

\begin{center}
\allowbreak FIGURE CAPTIONS

\bigskip

\bigskip

\bigskip
\end{center}

Fig. 1 - Behaviors of the nondimensional temperature and heat flux of
Brownian motion equation (MBE), Equation of phonon radiative transfer (EPRT)
and ballistic diffusive equations (BDE) for $Kn=1$ and $t^{\ast }=1$. The
data of the EPRT and BDE have been extracted from Ref. \cite{chen}.

\bigskip

Fig. 2 - Plot of the ratio $\Lambda =-\phi /(\partial _{\xi }\theta )$ in
function of the nondimensional coordinate $\xi $ calculated from the data of
EPRT (Fig.1).

\bigskip

Fig. 3 - Comparison \ of the nondimensional temperature and heat flux in
terms of \ the nondimensional coordinate $\xi $ obtained from the EPRT, FPE
and BDE for $Kn=1$ and $t^{\ast }=1.$ In the case of FPE, the results have
been calculated by using $a(t^{\ast })=$\ $(1-e^{-ht^{\ast }})$ and $%
D_{1}(\xi )=\left( 1+p_{2}\xi ^{4}\right) /\left( 1+p_{3}\xi ^{4}\right) $
with the parameters given by $h=1.01$,  $p_{2}=0.7$ and $p_{3}=2.4$.

\bigskip

Fig. 4 - Comparison \ of the nondimensional temperature and heat flux in
terms of the nondimensional coordinate $\xi $ obtained from the EPRT, FPE
and BDE for $Kn=1$ and $t^{\ast }=1.$ In the case of FPE, the results have
been calculated by using $a(t^{\ast })=$\ $(1-e^{-ht^{\ast }})$ and $%
D_{2}(\xi )=p_{3}\left( 1+p_{2}\xi ^{4}\right) /[p_{2}\left( 1+p_{3}\xi
^{4}\right) ]$ with the parameters given by $h=0.265$, $p_{2}=0.63$ and $%
p_{3}=1.015$.

\bigskip

Fig. 5 - Comparison of the nondimensional heat flux in function of the
nondimensional time $t^{\ast }$ obtained from the BME and FPE at $\xi =0$.
In this plot we have used $a(t^{\ast })=$\ $(1-e^{-ht^{\ast }})$ and $%
D_{1}(\xi )=\left( 1+p_{2}\xi ^{4}\right) /\left( 1+p_{3}\xi ^{4}\right) $
with the parameters given by $h=1.01$, $p_{2}=0.7$ and $p_{3}=2.4.$ The
solid lines correspond to the BME data, whereas the dotted lines correspond
to the FPE data. The pairs of the curves from top to bottom are calculated
by using $Kn=10,1,0.1$, respectively.

\end{document}